# Effective stress law for the permeability of a limestone

Siavash Ghabezloo[1,*], Jean Sulem[1], Sylvine Guédon[2], François Martineau[2]

[1]Université Paris-Est, UR Navier, CERMES, Ecole Nationale des Ponts et Chaussées, 6-8 avenue Blaise Pascal, Cité Descartes, 77455 Champs-sur-Marne, Marne la Vallée cedex 2, France

[2]Université Paris-Est, LCPC, MSRGI, 58, boulevard Lefebvre, 75015 Paris Cedex 15, France

## Abstract

The effective stress law for the permeability of a limestone is studied experimentally by performing constant head permeability tests in a triaxial cell with different conditions of confining pressure $\sigma$ and pore pressure $p_f$. Test results have shown that a pore pressure increase and a confining pressure decrease both result in an increase of the permeability, and that the effect of the pore pressure change on the variation of the permeability is more important than the effect of a change of the confining pressure. A power law is proposed for the variation of the permeability with the effective stress ($\sigma' = \sigma - n_k p_f$). The permeability effective stress coefficient $n_k$ increases linearly with the differential pressure and is greater than one as soon the differential pressure exceeds few bars. The test results are well reproduced using the proposed permeability-effective stress law. A conceptual pore-shell model based on a detailed observation of the microstructure of the studied limestone is proposed. This model is able to explain the experimental observations on the effect of the total stress and of the pore pressure on the permeability of the limestone. Effective stress coefficients for the stress-dependent permeability which are greater than one are obtained. It is shown that the controlling factor is the ratio of the different bulk moduli of the various constituents of the rock. This ratio is studied experimentally by performing microhardness tests.

---

[*] Corresponding Author, Email: ghabezlo@cermes.enpc.fr

# 1  Introduction

The determination of the permeability of a rock and its variation with changes of total stress and pore pressure has a great interest in petroleum reservoir engineering. The pore pressure decrease during the oil or gas production from a reservoir causes the reduction of the permeability of the reservoir and consequently the reduction of the production rate. The variations of permeability with the state of stress is also important in geophysics in the studies of rapid fault slip events when shear heating tends to increase the pore pressure and to decrease the effective compressive stress and the shearing resistance of the fault material (Rempel and Rice [1], Sulem et al. [2]). The increase of the permeability of the fault material with the pore pressure increase can lead to a faster dissipation of the excess pore pressure and consequently can decelerate the weakening of the shearing resistance of the fault material.

If a single variable can be defined as a linear combination of the total stress and the pore pressure to express the variations of the permeability, then we can say that the permeability follows an effective stress law. The concept of effective stress was first introduced by Terzaghi [3] who defined it as the difference between the total stress $\sigma$ and pore pressure $p_f$ and attributed all measurable effects of a change in stress exclusively to changes in the effective stress:

$$\sigma_d = \sigma - p_f \tag{1}$$

In equation (1) $\sigma_d$ is the Terzaghi effective stress, which is also called the differential pressure. More generally, the effective stress $\sigma' = \sigma'(\sigma, p_f)$ can be defined as a stress quantity which can be used as a single variable to express the stress dependency of a property $Q$ of a porous material (Equation (2)). Reducing the number of independent variables from two to one by using the concept of effective stress greatly simplifies the analysis of total stress and pore pressure dependency of the porous material properties.

$$Q = Q(\sigma, p_f) = Q(\sigma') \tag{2}$$

If the total stress and the pore pressure vary in such a way that the effective stress remains constant, then no variation in the corresponding property $Q$ is expected. Thus the expression of the effective stress can be obtained from the evaluation of the isolines of $Q(\sigma, p_f)$.

Since different material properties may depend on total stress and pore pressure in different ways, there is not a unique effective stress which would be appropriate for all properties of the material, and consequently different effective stress expressions should be defined for the different properties.

Assuming that the function $Q(\sigma, p_f)$ is smooth enough so that its derivatives can be defined, the incremental variation of $Q$ can be written as:

$$dQ = \frac{\partial Q}{\partial \sigma} d\sigma + \frac{\partial Q}{\partial p_f} dp_f \tag{3}$$

We can re-write equation (3) in the following form:

$$dQ = \frac{\partial Q}{\partial \sigma}\left[d\sigma - \left(-\frac{\partial Q/\partial p_f}{\partial Q/\partial \sigma}\right)dp_f\right] \tag{4}$$

The above expression shows that the variation of the property $Q$ can be expressed as a function of a single incremental quantity $d\sigma'$:

$$dQ = \frac{\partial Q}{\partial \sigma} d\sigma' \tag{5}$$

with

$$d\sigma' = d\sigma - n_Q dp_f \tag{6}$$

The coefficient $n_Q$ is the effective stress coefficient corresponding to the property $Q$ defined as:

$$n_Q(\sigma, p_f) = -\frac{\partial Q/\partial p_f}{\partial Q/\partial \sigma} \tag{7}$$

The isolines of $Q$ are obtained by the integration of the following differential equation:

$$d\sigma' = 0 \tag{8}$$

In the close vicinity of a given state of stress ($\sigma, p_f$), the isolines are generally approximated with parallel straight lines (e.g. Bernabé [4]) which is equivalent to the assumption that $n_Q$ is a constant. Under this assumption, equation (6) can be easily integrated and a linear expression for the effective stress is obtained:

$$\sigma' = \sigma - n_Q p_f \tag{9}$$

The linear expression presented in equation (9) is the most common expression for the effective stress as it is usually used in the mechanics of porous media. The effective stress coefficient $n_Q$ is equal to one in Terzaghi's definition which means that the total stress and the pore pressure have similar, but inverse, effects on the variation of the property $Q$. A value of the coefficient $n_Q$ smaller (respectively greater) than unity means that the effect of pore pressure change on the variation of the property $Q$ is less (respectively more) important than the effect of a change in total stress.

Zimmerman [5] and Berryman [6] have derived general effective stress rules for various physical properties of rocks and presented the expressions of the effective stress coefficients $n_Q$ corresponding to different physical properties together with some bounds and general relations among these coefficients.

As mentioned earlier, the great advantage of the effective stress concept is the simplification of the analysis by using one single variable $\sigma'$ instead of two independent variables, $\sigma$ and $p_f$. The resulted simplification is significant when the effective stress coefficient is constant, but the variability of this coefficient adds new difficulties to the problem. In this case, if the effective stress coefficient is a simple function of the pore pressure and the confining pressure, for example a linear function of the differential pressure $\sigma_d$, then the use of the effective stress concept may be still advantageous. Obviously, if the effective stress coefficient involves a complicated relation between the pore pressure and the confining pressure, it may be more appropriate to abandon the effective stress concept and to analyse the problem using the two independent variables.

The effective stress corresponding to the variations of the permeability of the rocks has been studied theoretically and experimentally by different researchers. Berryman [6] has studied the problem theoretically and has derived the expressions of the effective stress coefficient for the permeability of the rocks constituted by one single and two different minerals. His approach resulted in an effective stress coefficient which is smaller than one for a micro-homogeneous material made up of one single mineral and which may be greater than one for a porous material constituted by two different minerals. This is for example the case in clay-rich sandstones as the grains modulus of the pore-filling material is much smaller than the one of the sand grains. Experimental studies on both sedimentary and crystalline rocks can be found in the literature. Most of the studies on sedimentary rocks have been conducted on clay-rich sandstones (Zoback [7], Zoback and Byerlee [8], Walls and Nur [9], Nur et al. [10], Walls [11], Al-Wardy [12], Al-Wardy and Zimmerman [13]) and resulted in effective stress coefficients which are greater than one and which increase with the clay content. The evaluated effective stress coefficient by Walls [11] varies from 1.2 for clean sandstone to 7.1 for sandstone containing 20% clay. These results show that the permeability of the studied rock is more sensitive to variations of the pore pressure than to the changes of the total stress. To explain these results, Zoback and Byerlee [8] have proposed a conceptual clay-shell model in which the rock consists of quartz, permeated with cylindrical pores that are lined with an annular layer of clay which is more compliant than the outer quartz layer. Such a model is more sensitive to changes in pore pressure than to changes in total stress, and therefore results in an effective stress coefficient greater than one. Al-Wardy and Zimmerman [13] have proposed a different microstructural model in which the clay is distributed in the form of particles that are weakly coupled to the pore walls and have obtained results that are more compatible with the experimental results. Kwon et al. [14] have studied the permeability of illite-rich shale with a clay content of about 45% and found an effective stress coefficient equal to one which differs from the results obtained for clay-bearing sandstones with low clay content. They concluded that this is due to the contiguity of clays throughout the rock which support the applied confining pressure as well as the pore pressure.

As opposed to the results obtained for porous sedimentary rocks, the experimental studies performed on crystalline and jointed rocks resulted in effective stress coefficients which are smaller than or very close to one. Walsh [15] examined the experimental results of Kranz et al. [16] for flow through joints with polished surfaces and joints with jagged surfaces formed by a tensile fracture. The estimated effective stress coefficients are equal to 0.9 for the polished surfaces and 0.56 for the tension fractures. Experimental studies of Coyner [17] and Morrow et al. [18] on two granites resulted in permeability effective stress coefficients very close to one. Bernabé [4],[19] studied the permeability of several crystalline rocks and found effective stress coefficients which are smaller than one and which decrease with increasing confining pressure due to the changes in the geometry of the cracks during closure. He also reported the strong loading path dependency of the effective stress coefficient. This stress path dependency decreases rapidly with the number of cycles and the effective stress coefficient approaches one after a few cycles.

In this paper an experimental study for determining the effective stress law for the permeability of a limestone is presented. A drained hydrostatic compression test is performed in order to evaluate the drained bulk modulus of the rock. Constant head permeability tests are performed with different conditions of confining pressure and pore pressure and the effective stress coefficient for the variations of the permeability of the rock is determined. The results enable us to compare the sensitivity of the permeability of the rock to the variations of the total stress and of the pore pressure. Microscopic observations have been performed in order to study the microstructure of the limestone and to better understand the mechanism of permeability evolution with total stress and pore pressure variations. On the basis of these observations, a conceptual model is proposed for the microstructure of the material in order to determine the effective stress coefficient for the permeability of the rock and to interpret the experimentally measured values of this coefficient. This model accounts for the relative compliance of the constituents of the rock which is studied by performing microhardness tests.

## 2   Description of the studied rock

The rock studied in this paper is an oolitic limestone of Neocomian age (–135 to –127 million years), collected near Nîmes in south of France. This white limestone is homogeneous at the scale of the quarry, at the scale of the blocs and also at the scale of the samples observed under a microscope.

The observations are performed using a polarizing optical microscope and an environmental scanning electron microscope (ESEM). These two observation methods are complementary for this study, as they provide two different magnification scales.

The ESEM is equipped by two types of detectors: A detector of secondary electrons (SE) and a detector of backscattered electrons (BSE). The SE detector results in images which emphasize the topographical contrasts on the observed sample and have a three-dimensional appearance. In these images, steep surfaces and edges tend to be brighter than flat surfaces. The BSE detector can detect the contrast between areas with different chemical compositions, especially when the average atomic number of the various regions is different. The brightness of the images taken by this detector tends to increase with the atomic number.

The studied limestone is entirely composed of calcite, with calcite crystals of different sizes depending on the period and the dynamics of the crystallization. An image using the optical microscope under polarized light is presented in Figure (1a). This microphotograph shows that the studied rock is constituted of quasi spherical oolites, with a size of 100 to 400 µm, cemented by a sparitic calcite. An other image taken with the ESEM using the SE detector is presented in Figure (1b) and shows the details of the microstructure and the interface between oolites and sparitic calcite cement. The geological formation of this limestone can be summarized in three steps:

- Formation of concentric oolites in a high-energy marine environment at high temperature, low depth waters; these oolites are constituted from micro-crystallized calcite (micrite)

- Sedimentation of oolitic sand and crystallization of a palissadic calcite ore around the oolites; this sedimentation has also a marine origin

- Cementation of the oolites by precipitation of crystalline calcite of sparite type

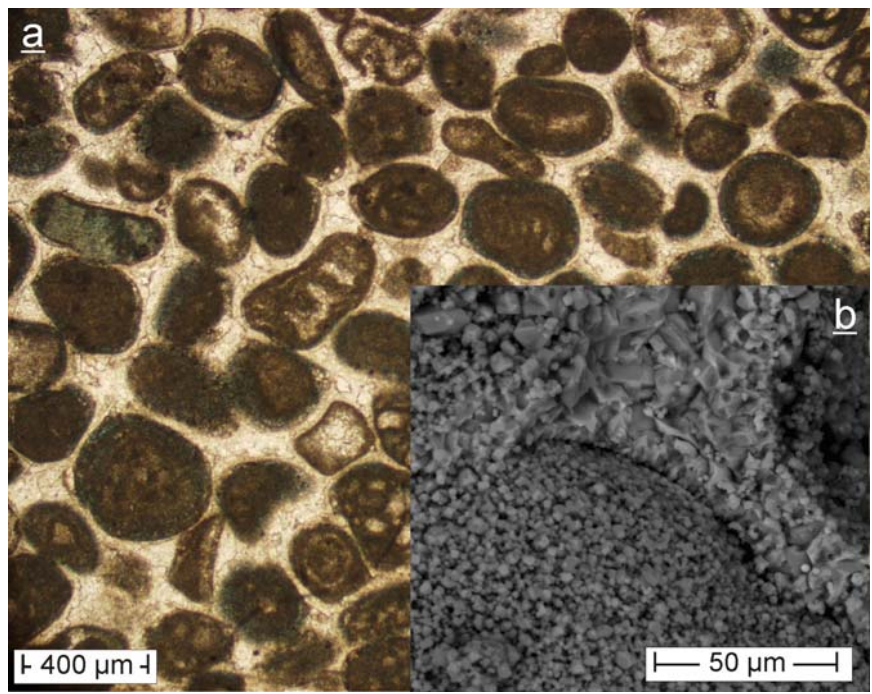

**Figure 1- Microstructure of the studied limestone observed with (a) optical microscope, (b) ESEM**

Figure (2a) presents the details of the micrite crystals forming the oolites. We can see that the microstructure of the oolites is similar to a granular material with grains diameters varying from 2 to 5 µm and with a fine porosity between the grains. The sparite crystals which constitute the cement connecting the oolites are shown in Figure (2b). We can see clearly the difference between the microstructure of the oolites and the one of the sparitic cement. This latter is composed of interlocked grains with a bigger size than the micrite grains of the oolites. Both sparite and micrite are entirely composed of calcite.

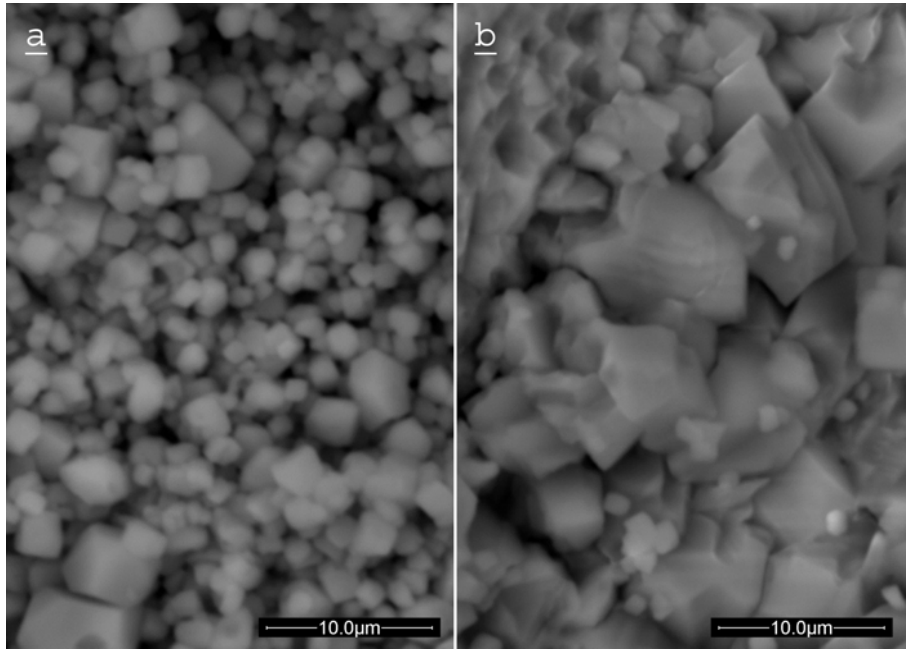

**Figure 2- Details of (a) an oolite and (b) the sparitic cement observed with the ESEM**

The measured connected porosity of the studied limestone is between 13.4% and 17.6% with the average value equal to 15.7%. The measured apparent density of the skeleton is equal to $2.22\,\text{g}/\text{cm}^3$ and the measured absolute density of the grains is equal to $2.70\,\text{g}/\text{cm}^3$ which is compatible with the value $2.71\,\text{g}/\text{cm}^3$ given by Bass [20] for calcite. From these values the total porosity of the rock is evaluated equal to 17.7%. The difference between the connected and the total porosity reveals the presence of 2.0% occluded porosity in the rock matrix.

The porosity of the rock is studied using the optical microscope on a sample which was injected with a blue resin (Figure (3a)). The blue areas, which correspond to the pores, are mostly concentrated around the oolites in a layer with a thickness of 10 to 20 µm. The porosity of the rock can also be studied using the ESEM images taken with the BSE detector. In these images the porous space appears as in black areas. An image is shown in Figure (3b) and shows clearly that the porosity is concentrated around the oolites. Comparing this porosity with the one observed in Figures (2a) and (2b) we can conclude that there are two types of porosity in the studied limestone: a macro-porosity and a micro-porosity. The macro-porosity as seen in Figure (3) is mostly concentrated around the oolites while the micro-porosity corresponds to the intergranular voids between the grains of micrite in the oolite and between the grains of sparite in the cement. The analysis of several images showed that the blue areas, which represent the macro-porosity, fill about 13% of the total area of the images. Comparing this value to the average connected porosity of the rock, 15.7%, one can deduce that most of the connected porosity is located around the oolites. Comparing the evaluated macro porosity, about 13%, and the total porosity, 17.7%, we can estimate the micro-porosity of the rock equal to 4.7%. As mentioned above, 2.0% of the porosity of the rock is occluded. Comparing this value with the micro-porosity, we can infer that about 42% of the micro-porosity of the rock is occluded.

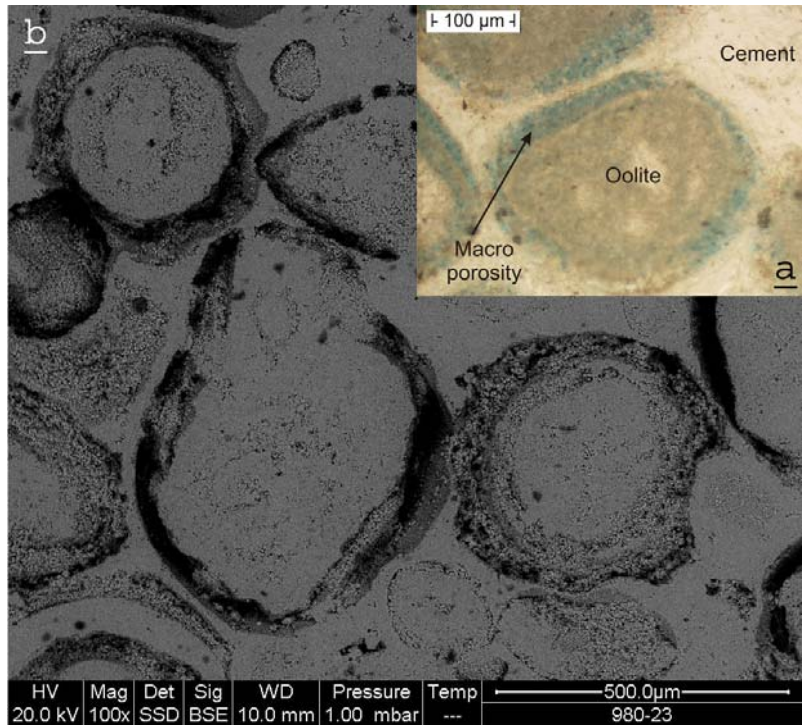

**Figure 3- Porous space of the limestone-concentration of the main porosity around the oolites, (a) optical microscope (b) ESEM**

The analysis of the image presented in Figure (1a) shows that about 28% of the area of this image is filled with the sparitic cement and thus the remaining 72% is filled with the oolites and the macro porosity. Knowing that the macro porosity is equal to 13% as presented above, the volumetric fraction of the oolites is found equal to 59%.

## 3   Laboratory tests

Laboratory tests in this study consist of a drained hydrostatic compression test for the evaluation of the compressibility of the rock and constant head permeability tests with different values of pore pressure and confining pressure in order to study their individual effect on the variations of the permeability. Note that all the tests are performed on the same sample. The drained compression test with a loading-unloading cycle is performed before the permeability tests. This guaranties that the sample behaves elastically during the permeability tests. The tests are performed on a 40mm diameter and 80mm length saturated sample in a triaxial cell as described in Ghabezloo and Sulem [21]. The saturation procedure is the following: The sample is first dried at a temperature of 60°C until the measured mass remains constant and then it is put in a vacuum chamber during 24 hours to eliminate the remaining air bubbles. Water is then introduced in the chamber and the sample is kept immerged under water during 48 hours. The sample is then put in a rubber jacket with a grease layer inside the jacket and installed inside the triaxial cell. An initial confining pressure is applied and water is circulated using a pressure generator.

## 3.1 Drained hydrostatic compression test

The bulk modulus of the rock is evaluated in a drained hydrostatic compression test with a loading and unloading cycle. In this test, a fluid back pressure of 1.0MPa was maintained constant inside the sample. In Figure (4) the volumetric strain response is shown versus the applied total stress. The observed non-linear response reflects the stress-dependent character of the rock compressibility (Zimmerman [5], Sulem et al. [22][23]). Moreover this curve exhibits a different response in loading and unloading and irreversible deformation. Therefore the elastic tangent drained compression modulus $K_d$ is evaluated on the unloading curve. The slope of this curve is directly evaluated of the experimental data using the method used in Ghabezloo and Sulem [21] where the slope of the tangent at each point is calculated as a difference quotient of *N* data points centred around the corresponding point. The evaluated bulk modulus is presented as a function of Terzaghi effective stress on Figure (5). This curve can be approximated by the following linear expression:

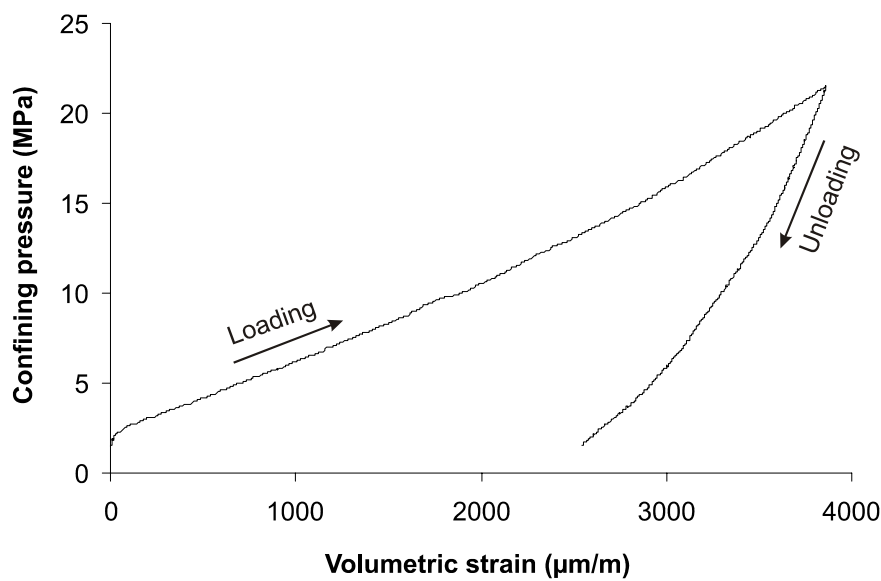

**Figure 4- Drained hydrostatic compression test with a loading-unloading cycle**

$$K_d = 6.22 + 1.035\sigma_d \qquad (K_d : \text{GPa}, \sigma_d : \text{MPa}) \qquad (10)$$

This expression shows the linear increase of the drained bulk modulus with Terzaghi effective stress in the limit of the applied pressures. The non linear elastic response in hydrostatic compression can be attributed to the closure of the regions of imperfect bounding between grains (Zimmerman [5]) and the compaction of the porous space of the rock. Figure (3b) exhibits limited contact areas between the oolites and the surrounding cement, which can cause the compressibility of the microstructure of the rock. More contacts can be formed under stress leading thus to a strengthening of the rock. This strengthening effect will be limited at higher stress due to saturation of the network of contact points and possible grains breakage. Thus the linear increase of the bulk modulus is only valid at low stress.

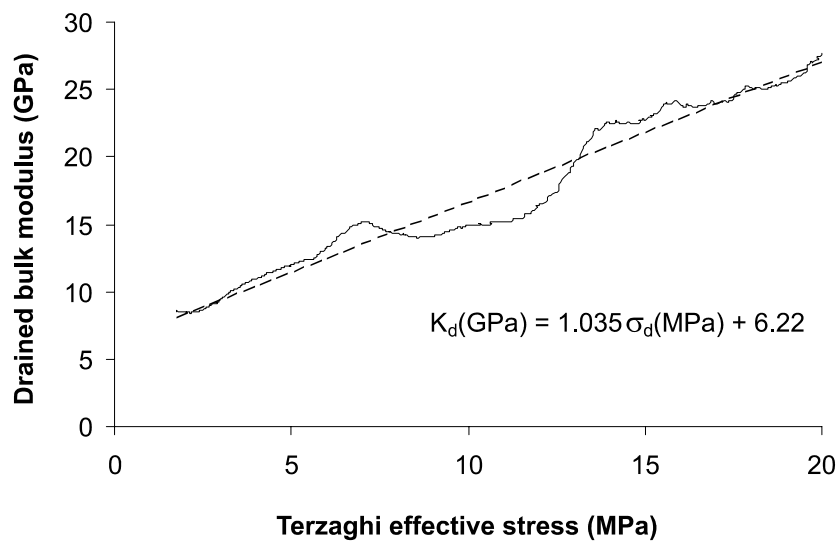

**Figure 5- Drained bulk modulus of the studied limestone as a function of Terzaghi effective stress**

## *3.2 Permeability tests*

Constant head permeability tests have been performed in a triaxial cell on a water saturated limestone sample under isotropic loading. After the saturation, two pore pressure generators were used to keep a constant pore pressure difference of 300 kPa between the two ends of the sample. The confining pressure was always kept about 2MPa higher than the pore pressure to prevent the fluid infiltration between the sample and the jacket. The permeability tests have been performed with different combinations of pore pressure and confining pressure. The mean value of the two pore pressures imposed at the ends of the sample is taken as the sample average pore pressure for the evaluation of the effective stress inside the sample. After the saturation of the sample under the initial confining pressure, the pore pressure at each end of the sample was maintained constant and the confining pressure was increased by steps of 1.0MPa. At each step the flow rate was recorded after the stabilization of the fluid flow and the permeability of the sample was calculated using the Darcy's law. This procedure was repeated for three different pore pressure levels with eight confining pressure steps for each level. Thus, 24 permeability tests have been performed with various combinations of pore pressure and confining pressure. The results, presented in Figure (6), show the measured permeability as a function of the confining pressure for different average pore pressures. These results show that the permeability decreases with the confining pressure increase for a constant average pore pressure. Moreover, under a constant confining pressure, the permeability increases with the pore pressure increase. These experimental results can be explained by the pore diameter changes under the effect of the confining pressure and the pore pressure variations. The shape of the permeability curves and its reduction with the confining pressure increase is similar to the results reported in the literature for porous sedimentary rocks (Coyner [17], Yale [24], David et al. [25], Schutjens et al. [26]) as well as for crystalline rocks (Kranz et al. [16], Coyner [17], Bernabé [19]). As mentioned by David et al. [25], the compaction mechanism in crystalline rocks is related to the closure of microcracks and the pressure sensitivity of permeability decreases with confining pressure increase. This pressure sensitivity of the permeability is more important in crystalline rocks than in porous sedimentary rocks for which the compaction is related to the relative movement of grains. Vairogs et al. [27] showed that the pressure sensitivity is more important for the rocks with lower permeability.

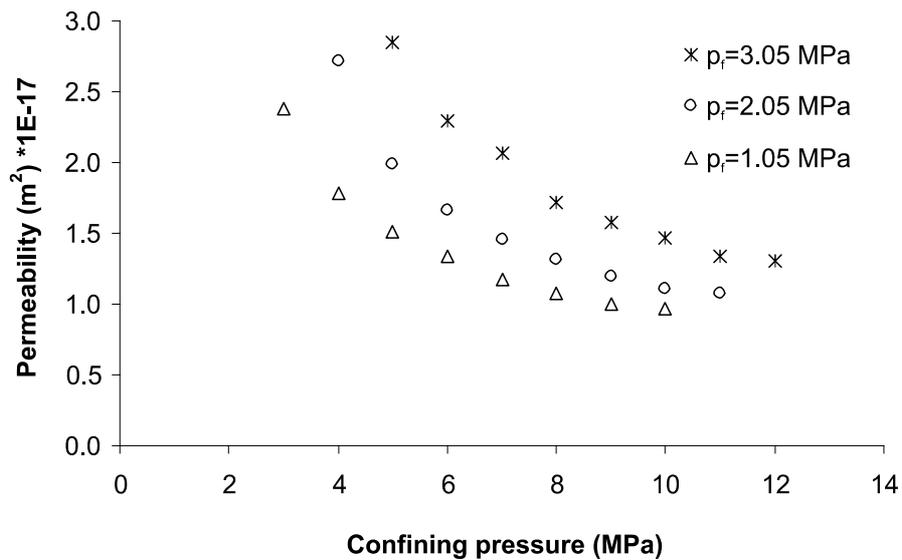

**Figure 6- Results of permeability tests performed with different conditions of pore pressure and confining**

On Figure (6), for the range of the applied confining pressures, the variation of the permeability with the confining pressure follows a power law. Even though this power law cannot be accepted as a general permeability-stress relationship due to the infinite value at zero stress, it can be accepted to express the permeability variations with the variations of the confining pressure in the limit of the applied pressures.

## 4 Analysis of test results

The easiest way to determine the effective stress coefficient for the variations of the permeability based on the obtained experimental results is to draw a contour plan of the permeability as a function of pore pressure and confining pressure. The slope of the isolines gives the effective stress coefficient corresponding to the variations of the permeability. Parallel isolines would mean that the effective stress coefficient is constant. The contour plan is shown in Figure (7). In this figure, the same scale is used for the pore pressure and the confining pressure axes so that the slope of the isolines can be calculated directly on the graph. We can clearly see that the isolines are not parallel and that the effective stress coefficient calculated as the average slope of the isolines decreases with increasing values of the permeability. The effective stress coefficients evaluated on this graph varies between 0.9 and 2.4. Considering the limited number of data points on Figure (7), it should be mentioned that most of the fine details of the isolines on this figure are just artefacts. Moreover, considering the small difference between the measured permeabilities at higher confining pressures, the evaluated effective stress coefficients should be seen as approximate values. Nevertheless, these data clearly show the general trend for the evolution of the effective stress coefficient as described above.

In order to represent the effect of total stress and pore pressure on the permeability and on the effective stress coefficient, a power law for the variation of the permeability with the effective stress as defined by equation (9) is assumed:

$$k = a\sigma'^b \tag{11}$$

The effective stress coefficient $n_k$ is assumed to vary linearly with the differential stress (i.e. Terzaghi effective stress):

$$n_k = c\sigma_d + d \tag{12}$$

The coefficients *a* and *b* in equation (11) and *c* and *d* in equation (12) are evaluated using a least square algorithm by fitting the experimental data. The following expressions for the effective stress coefficient and for the permeability-effective stress dependency are found:

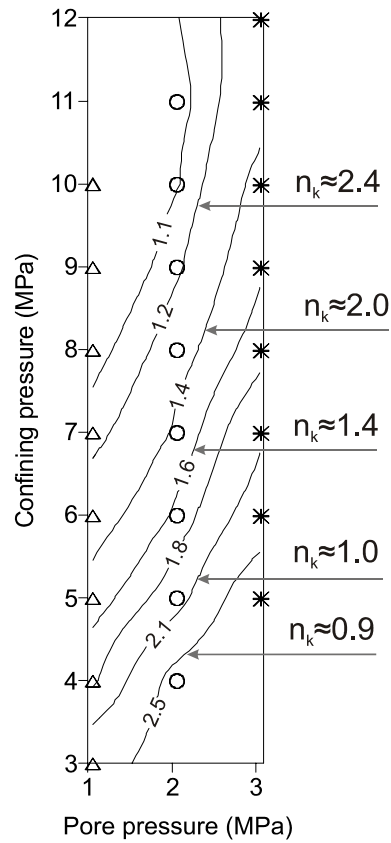

**Figure 7- Permeability contour plan and the approximate effective stress coefficient corresponding to each isoline**

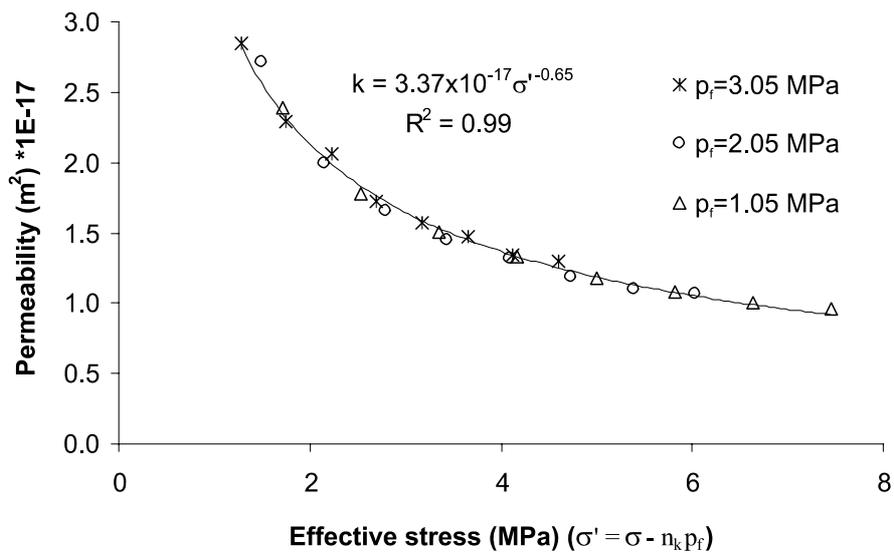

**Figure 8- Relation between permeability and effective stress**

$$n_k = 0.89 + 0.17\sigma_d \qquad (\sigma_d : \text{MPa}) \tag{13}$$

$$\sigma' = \sigma - n_k p_f \tag{14}$$

$$k = 3.37 \times 10^{-17} \sigma'^{-0.65} \qquad \left(k : \text{m}^2, \sigma': \text{MPa}\right) \tag{15}$$

The results of the curve fitting procedure are shown on Figure (8). We can observe that the experimental permeability measurements are well superposed when plotted against the obtained effective stress. Figure (9) shows the simulation of the permeability tests using the above permeability law. This graphs show that the experimental data are well reproduced and that the contour plans of the measured and computed permeability are very well superposed.

As mentioned earlier, the main advantage of the effective stress concept is its simplicity. Thus, one could simplify the stress-dependent permeability law given by equations (13) and (15) by neglecting the variations of the effective stress coefficient and take a mean value for $n_k$, equal to 1.3.

An effective stress coefficient which is greater than unity means that the effect of pore pressure variation on the permeability of the limestone is more important than the effect of the variation of the total stress. This result is similar to the ones presented by Zoback and Byerlee [8], Walls and Nur [9], Walls [11] and Al-Wardy and Zimmerman [13] for clay-rich sandstone. In the clay-shell and clay-particle models proposed by Zoback and Byerlee [8] and Al-Wardy and Zimmerman [13], an effective stress coefficient greater than one is interpreted by the presence of the clay in the rock and the great difference between the elastic moduli of the clay and the quartz. For the studied limestone a simple conceptual model for the microstructure is proposed in the following section in order to understand the mechanism of the variation of the permeability and of the corresponding effective stress coefficient with the confining pressure and the pore pressure.

## 5  Pore-shell model

Based on the microscopic observations of the limestone, a simplified model for the microstructure of the studied rock is proposed in this section. This model is presented in Figure (10) and consists of a central oolite circular core that is covered by a pore shell and in the external part by a cement shell. Following the observation that an important portion of the connected porosity of the studied limestone is concentrated around the oolites, it is assumed that the porosity of the rock is entirely situated around the central oolite core. Consequently the micro-porosity of the rock observed between the micrite and sparite grains is neglected and oolite and cement are considered as non-porous elastic materials. This assumption is supported by the fact that about 42% of the micro-porosity of the rock, situated in the oolites and in the sparitic cement, is occluded. This model is in its principles similar to the clay-shell model and to the clay-particle model proposed respectively by Zoback and Byerlee [8] and Al-Wardy and Zimmerman [13]. Similarly to the assumption of non porous cement and oolites in the proposed pore-shell model, a mechanism is needed to prevent the penetration and distribution of the pore pressure inside the clay phase in the clay-shell and clay-particle models. Indeed, as mentioned by Coyner [17], the compressibility of clay particles and of the rock solid grains are comparable so that if the pore pressure freely infiltrates into the clay pore space, the resulting effect will be equivalent of an 'unjacketed' compression test and will lead to a very small deformation caused by the compressibility of clay particles, and consequently the permeability change will be much less sensitive to the pore pressure variations.

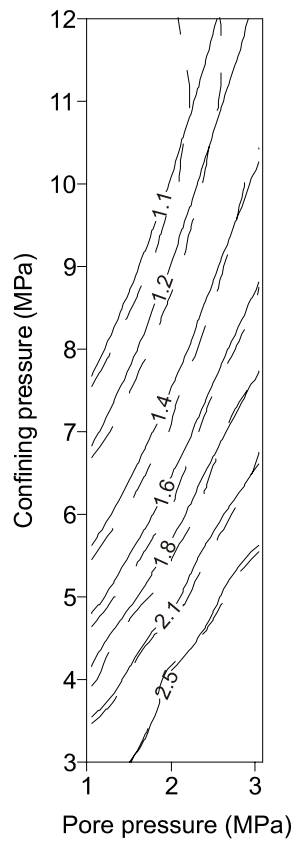

**Figure 9- Comparison of the contour plans of the experimental results (dashed lines) and the proposed permeability-effective stress model (solid lines)**

The oolites in the studied limestone are quasi-spherical. Moreover, as can be seen in Figure (3), the macro-porosity situated around the oolites is interconnected. Therefore we can consider a spherical geometry for the proposed pore-shell model. The problem can also be solved using a cylindrical geometry, as it is derived for the clay-shell and clay-particle models. In order to compare the results obtained with these two geometries, the solution for cylindrical geometry is presented in Appendix 1. As shown by Al-Wardy [12], it is expected that the results obtained from cylindrical and spherical geometries do not differ much.

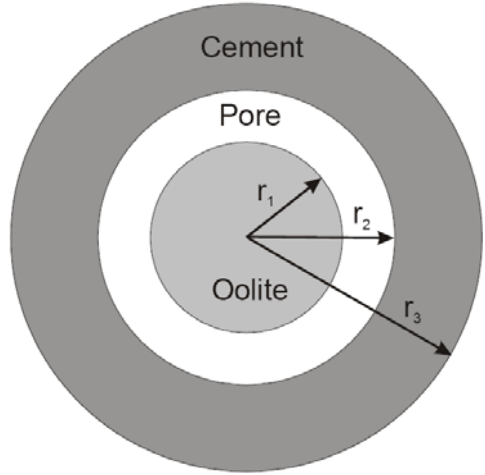

**Figure 10- Pore-Shell model**

## *5.1 Model formulation*

The permeability of the proposed model is a function of the pore-shell geometry and thus a function of the pore-shell thickness:

$$k = f(r_2 - r_1) \qquad (16)$$

Assuming the permeability $k$ as the property $Q$ in equation (7) and using equation (16), the expression of the effective stress coefficient $n_k$ for the permeability of the model is written as:

$$n_k = -\frac{\left(\frac{\partial (r_2 - r_1)}{\partial p_f}\right)_\sigma}{\left(\frac{\partial (r_2 - r_1)}{\partial \sigma}\right)_{p_f}} \qquad (17)$$

The radius of the central oolite core ($r_1$) is independent of the confining pressure $\sigma$, thus the following expression is obtained for $n_k$:

$$\frac{\partial r_1}{\partial \sigma} = 0 \quad \Rightarrow \quad n_k = \frac{-\frac{\partial r_2}{\partial p_f} + \frac{\partial r_1}{\partial p_f}}{\frac{\partial r_2}{\partial \sigma}} \qquad (18)$$

The confining pressure is applied on the external boundary of the model and the pore pressure is applied inside the pore-shell. The elastic solution of the problem and the variations of the model geometry parameters with the applied pressures are obtained from the solution of a hollow sphere problem.

Considering a hollow sphere with the inner radius $a$ and the outer radius $b$, the radial stresses at the inner and outer boundaries are respectively $-p_i$ and $-p_o$. Compressive stresses are taken negative. The well-known Lamé solution holds for the radial displacement $u(r)$:

$$u(r) = Ar + \frac{B}{r^2} \tag{19}$$

The integration constants $A$ and $B$ are obtained from the boundary conditions:

$$A = \frac{p_i a^3 - p_o b^3}{(3\lambda + 2\mu)(b^3 - a^3)} \tag{20}$$

$$B = \frac{(p_i - p_o) a^3 b^3}{4\mu(b^3 - a^3)} \tag{21}$$

where $\lambda$ and $\mu$ are the Lamé coefficients of the elastic cylinder.

For the central oolite core $a = 0$, $b = r_1$ and $p_o = p_f$ and let's denote by $\lambda_o$ and $\mu_o$ the Lamé coefficients of the oolite core. Using equations (19), (20) and (21) the expression of the displacement is found as:

$$u(r) = \frac{-p_f}{3\lambda_o + 2\mu_o} r \tag{22}$$

The term $\partial r_1 / \partial p_f$ in equation (18) is thus given by:

$$\frac{\partial r_1}{\partial p_f} = \frac{-r_1}{3\lambda_o + 2\mu_o} \tag{23}$$

For the outer cement shell $a = r_2$, $b = r_3$, $p_i = p_f$ and $p_o = \sigma$ and let's denote by $\lambda_c$ and $\mu_c$ the Lamé coefficients of the cement shell. Using equations (19), (20) and (21) the expression of the displacement is found as:

$$u(r) = \frac{p_f r_2^3 - \sigma r_3^3}{(3\lambda_c + 2\mu_c)(r_3^3 - r_2^3)} r + \frac{(p_f - \sigma) r_2^3 r_3^3}{4\mu_c (r_3^3 - r_2^3) r^2} \tag{24}$$

The terms $\partial r_2/\partial p_f$ and $\partial r_2/\partial \sigma$ in equations (18) are given by:

$$\frac{\partial r_2}{\partial p_f} = \frac{r_2^4}{(3\lambda_c + 2\mu_c)(r_3^3 - r_2^3)} + \frac{r_2 r_3^3}{4\mu_c (r_3^3 - r_2^3)} \tag{25}$$

$$\frac{\partial r_2}{\partial \sigma} = -\frac{r_2 r_3^3}{(3\lambda_c + 2\mu_c)(r_3^3 - r_2^3)} - \frac{r_2 r_3^3}{4\mu_c (r_3^4 - r_2^4)} \tag{26}$$

Introducing (23), (25) and (26) in equation (18) with some algebraic re-arrangements, the following expression is obtained for the effective stress coefficient corresponding to the variations of the permeability of the model:

$$n_k = 1 + \frac{4\mu_c \left(1 - (r_2/r_3)^3\right)}{3\lambda_c + 6\mu_c} \left( \frac{(3\lambda_c + 2\mu_c) r_1}{(3\lambda_o + 2\mu_o) r_2} - 1 \right) \tag{27}$$

Equation (27) can be re-written using the bulk modulus $K$ and of the Poisson's ratio $\nu$ ($\lambda = \frac{3K\nu}{1+\nu}$, $\mu = \frac{3K(1-2\nu)}{2(1+\nu)}$) of the material:

$$n_k = 1 + \left(1 - (r_2/r_3)^3\right) \frac{2(1-2\nu_c)}{3(1-\nu_c)} \left( \frac{K_c r_1}{K_o r_2} - 1 \right) \tag{28}$$

Note that only the Poisson's ratio of the cement appears in the above relation, whereas the Poisson's ratio of both the oolites and the cement appear for the cylindrical geometry as shown in equation (40) in Appendix 1.

The range of variations of the Poisson's ratio of rocks is relatively narrow. In order to simplify equation (28), we assume that $\nu_c = 0.3$ which is the value of the Poisson's ratio of calcite presented by Bass [20]. Using this assumption equation (28) is reduced to the following form:

$$n_k = 1 + \frac{8}{21} \left(1 - (r_2/r_3)^3\right) \left( \frac{K_c r_1}{K_o r_2} - 1 \right) \tag{29}$$

Based on the microscopic observations and the porosity measurements presented in the previous section, the following values are considered for the geometry of the model:

$$r_1 = 125\,\mu\text{m},\ r_2 = 135\,\mu\text{m},\ r_3 = 150\,\mu\text{m} \tag{30}$$

With these dimensions, the porosity of the model is equal to 15.0% which is approximately equal to the average measured connected porosity of the studied limestone, equal to 15.7%. Moreover, the volumetric fractions of different constituents of the microstructure of the rock, evaluated by image analysis and presented in the previous section, 59% oolite and 28% cement, are approximately retrieved with these values. In doing so, the following simple linear variation of the effective stress coefficient with the $K_c/K_o$ ratio is obtained:

$$n_k = 0.9 + 0.1 \frac{K_c}{K_o} \quad \text{(spherical geometry)} \tag{31}$$

In case of a microscopically homogeneous material, $K_c = K_o$, the coefficient $n_k$ is found equal to one. Note that for the cylindrical geometry, as presented in Appendix 1, we obtain a very close expression for the effective stress coefficient:

$$n_k = 0.92 + 0.07\, K_c/K_o \quad \text{(cylindrical geometry)} \tag{32}$$

We can see that the proposed pore-shell model, which is based on a detailed study of the microstructure of the limestone, is able to produce permeability effective stress coefficients that are greater than one and vary with the relative bulk moduli of the different constituents of the rock. The observations of the microstructure of the cement and of the oolites have shown that the interlocking of the particles in the cement can indeed lead to higher bulk modulus than for the oolites ($K_c/K_o > 1$). The $K_c/K_o$ ratio will be studied experimentally in the following section by performing microhardness tests.

## 5.2 *Microhardness testing of the limestone*

In the previous section it was shown that the permeability effective stress coefficient obtained from the proposed pore-shell model is a linear function of the $K_c/K_o$ ratio. The mechanical properties of the different constituents of the microstructure of the studied rock (mainly oolites and sparitic cement) are studied experimentally by performing Vickers microhardness tests. The Vickers hardness test consists of indenting the material with a diamond indenter, in the form of a right pyramid with a square base and an angle of 136° between opposite faces. The average length $d$ of the two diagonals of the indentation left on the surface of the material is measured. Obviously this length is smaller for a harder material. The Vickers hardness, $H_V$, is determined by the ratio of the load $p$ over the surface area of indentation and is calculated using the following simple equation:

$$H_v = 2\sin(136°/2)\frac{p}{d^2} = 1.8544\frac{p}{d^2} \tag{33}$$

The microhardness tests are performed on the oolites and cements separately and the tested sample is observed using an ESEM with the backscattered electron detector. From equation (33), the ratio of the hardness $H_{Vc}$ of the cement to the hardness $H_{Vo}$ of the oolites is given by:

$$\frac{H_{Vc}}{H_{Vo}} = \left(\frac{d_c}{d_o}\right)^{-2} \tag{34}$$

where $d_c$ and $d_o$ are the diagonals of the indentation on the cement and on the oolite respectively.

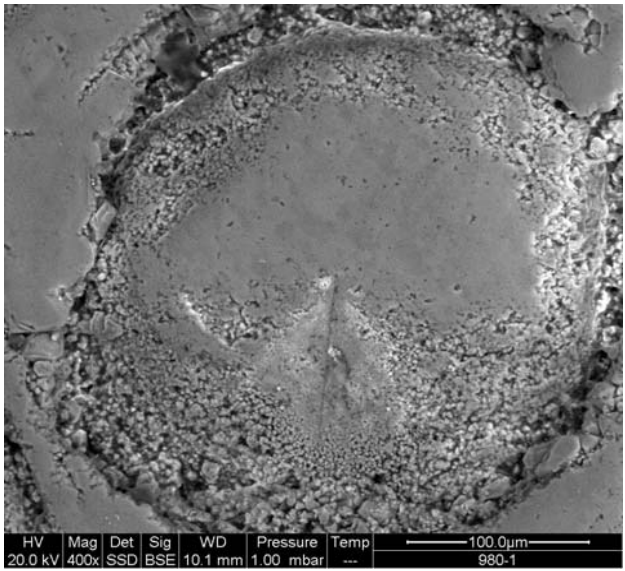
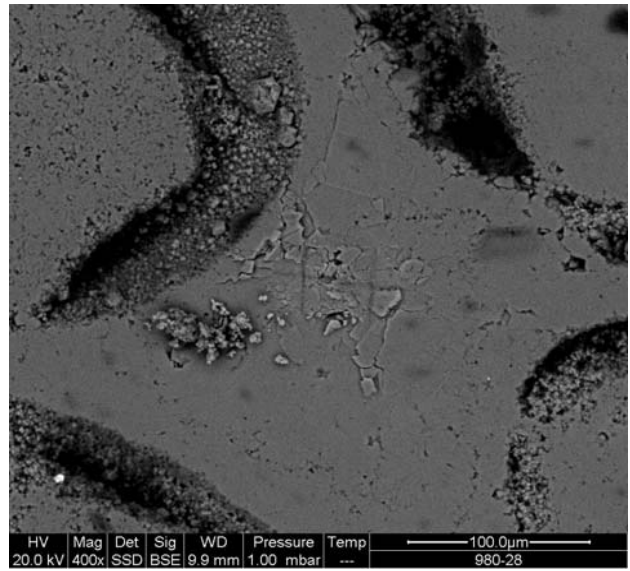

**Figure 11- Microhardness test: view of an indentation situated on an oolite**

**Figure 12- Microhardness test: view of two adjacent indentations situated on the sparitic cement**

The Vickers hardness is defined with reference to the area of the permanent impression (Das [28]), so it measures in some lumped way a combination of elastic, plastic, and fracture properties. There is, however, no unique relation between the hardness number and the primary parameters, but empirical relations may be obtained for certain classes of materials. For example, for a wide range of metals, the ratio of the hardness to uniaxial yield strength is a constant value between 2.7 and 3.0, but this relation does not hold for frictional materials, for which the hardness is a function of at least two variables: the cohesion and the friction angle of the material (Ganneau [29]). Only few empirical relations can be found in the literature between the Vickers hardness and the macroscopic elastic and strength properties of rocks. Das and Hucka [30] performed penetration tests on coal and showed a linear relation between the penetration strength and penetration modulus. They also presented a result from Honda and Sanada [31] which shows that the ratio of the Vickers hardness of coal to its Young's modulus is a constant. In absence of such data for limestone, we assume that the same type of relation holds for the constituents of the studied rock. Assuming also that the Poisson's ratio of the oolite and of the cement are equal, we obtain $K_c/K_o = H_{Vc}/H_{Vo}$.

The observation of the indentations of several performed microhardness tests resulted in a $H_{Vc}/H_{Vo}$ ratio which is greater than one and varies between 1.6 and 4.0. Figure (11) shows an image of an indentation situated on an oolite. Figure (12) shows two adjacent indentations situated on the sparitic cement. Both images are taken with the same magnification, so that the size of indented zones can be directly compared. This comparison clearly shows that the indentation on the oolite has a greater area and consequently that the hardness of the oolite is smaller than the hardness of the cement.

## 5.3 Discussion of model results

Inserting the values of $K_c/K_o$ ratio evaluated by microhardness tests in equation (31), we obtain that the effective stress coefficient is between 1.0 and 1.2. This is compatible with the result obtained using equation (13) for small differential pressures.

It should be noted that the $K_c/K_o$ ratio is obtained in the microhardness tests for zero confining pressure. It is expected that the bulk moduli $K_c$ and $K_o$ increase with the confining pressure due to the compaction of the pore space. This increase of the bulk moduli should be more important for the cement due to the interlocking of the particles and the greater coordination number than for the oolites. The ratio $K_c/K_o$ is thus expected to increase with the confining pressure which results in an increase of the permeability effective stress coefficient (equation (31)). In the absence of experimental data on the stress dependency of the bulk moduli $K_c$ and $K_o$, we can try to have a rough estimation from the drained bulk modulus of the rock (equation (10)) and considerations from homogenization theory. It is reasonable to assume that only the bulk modulus $K_c$ of the cement increases with confining pressure whereas the bulk modulus $K_o$ of the oolites remains constant. In the proposed pore-shell model, the confining pressure is applied only on the outer cement shell and has no effect on the oolite core. This may be realistic to some extent for low confining pressures. As can be seen in Figure (3), the oolites are connected to the surrounding cement with a few contact areas. The Voigt upper bound of a homogenized medium constituted by spherical oolites surrounded in a matrix of sparitic cement is given by:

$$K_d \leq f_o K_o + f_c K_c \tag{35}$$

where $f_o$ and $f_c$ are the volume fractions of the oolites and of the cement ($f_o + f_c = 1$). Considering the macro-porosity of the rock as a part of the oolite fraction, we get $f_o = 0.72$ and $f_c = 0.28$. For $\sigma_d = 0$ (unloaded state), equation (10) gives $K_d = 6.22\,\text{GPa}$ and for $\sigma_d = 8.95\,\text{MPa}$ (maximum applied differential pressure), $K_d = 15.5\,\text{GPa}$. From the results of microhardness tests we can take an average value for the ratio $K_c/K_o \simeq 3$ corresponding to the unloaded state. From equation (35) we obtain an upper bound for $K_o$ in the unloaded state equal to 4GPa. Keeping this value constant, the variation of $K_c$ is found from equation (35) to be between 12 and 34 GPa when $\sigma_d$ increases from 0 to 8.95MPa. Using equation (31), this results in an effective stress coefficient $n_k$ which varies between 1.2 and 2.0. This is comparable to the experimental results where $n_k$ was found to vary between 0.9 and 2.4.

It should be emphasized that this simple pore-shell model for the studied limestone is based on strong assumptions and is only used here for a better understanding of the mechanism of the evolution of the permeability with the pore pressure and the confining pressure without trying to obtain quantitative predictions.

# 6 Conclusion

The effective stress law for the permeability of a limestone is studied in an approach which combines laboratory testing, microstructural observations and modelling. 24 constant head permeability tests have been performed in a triaxial cell with various conditions of confining pressure and pore pressure. The experimental results have shown that the pore pressure increase and the confining pressure decrease both result in an increase of the permeability, but the effect of the pore pressure change on the variation of the permeability is more important that the effect of a change in the confining pressure. It is shown that an effective stress law can be defined for the permeability and that the corresponding effective stress coefficient increases linearly with the differential pressure and is greater than one as soon the differential pressure exceeds few bars. A power law is proposed for the variation of the permeability with the effective stress. The test results are well reproduced using the proposed permeability-effective stress law.

The microscopic observations of the microstructure of the rock shows that the studied limestone is entirely composed of calcite and is constituted of quasi spherical oolites cemented by sparitic calcite. The observations have shown that the main porosity of the rock is concentrated around the oolite cores. Based on the microscopic observations, a conceptual pore-shell model is proposed to study the effective stress law for the permeability of the studied limestone. The proposed model is able to reproduce permeability effective stress coefficients that are greater than one and which vary with the ratio of the bulk moduli of the two constituents of the rock. The ratio of the bulk moduli of the sparitic cement to the one of the oolites is studied experimentally by performing microhardness tests and is found to be greater than one. According to the proposed pore-shell model, this results in an effective stress coefficient which is greater than one. The microstructural approach is thus consistent with the experimental results of the permeability tests.

# 7 Appendix 1

In this appendix we present the equations of the pore-shell model for cylindrical plane strain geometry. All of the parameters are the same as the ones used in section 5.1. Considering a hollow cylinder with the inner radius $a$ and the outer radius $b$, and the radial stresses $-p_i$ and $-p_o$ respectively at the inner and outer boundaries, the radial displacement $u(r)$ is given by the following equation:

$$u(r) = Ar + \frac{B}{r} \tag{36}$$

With the integration constants $A$ and $B$ given by:

$$A = \frac{p_i a^2 - p_o b^2}{2(\lambda + \mu)(b^2 - a^2)} \tag{37}$$

$$B = \frac{(p_i - p_o)a^2 b^2}{2\mu(b^2 - a^2)} \tag{38}$$

Following the same procedure presented in section 5.1 for spherical geometry, the effective stress coefficient for the permeability of the cylindrical geometry is obtained as:

*Ghabezloo et al.: Effective stress law for the permeability of a limestone*$$n_k = 1 + \frac{\mu_c\left(1-(r_2/r_3)^2\right)}{\lambda_c + 2\mu_c}\left(\frac{(\lambda_c+\mu_c)r_1}{(\lambda_o+\mu_o)r_2}-1\right) \tag{39}$$

Equation (39) can be re-written using the bulk modulus $K$ and of the Poisson's ratio $v$ of the material:

$$n_k = 1 + \left(1-(r_2/r_3)^2\right)\frac{(1-2v_c)}{2(1-v_c)}\left(\frac{(K_c/(1+v_c))r_1}{(K_o/(1+v_o))r_2}-1\right) \tag{40}$$

Taking $v_c = v_o = 0.3$, equation (40) is reduced to the following form:

$$n_k = 1 + \frac{2}{7}\left(1-(r_2/r_3)^2\right)\left(\frac{K_c r_1}{K_o r_2}-1\right) \tag{41}$$

Based on the microscopic observations and the porosity measurements presented in the previous section, the following values are considered for the geometry of the cylindrical model:

$$r_1 = 125\,\mu\text{m},\ r_2 = 141\,\mu\text{m},\ r_3 = 165\,\mu\text{m} \tag{42}$$

With these dimensions, the porosity of the model is equal to 15.6% which is approximately equal to the average measured connected porosity of the studied limestone, equal to 15.7%. Moreover, the volumetric fractions of different constituents of the microstructure of the rock, evaluated by image analysis are approximately retrieved with these values. Inserting these dimensions in equation (41), the following linear variation of the effective stress coefficient with the $K_c/K_o$ ratio is obtained:

$$n_k = 0.92 + 0.07\frac{K_c}{K_o} \tag{43}$$